\documentclass[conference]{IEEEtran}

\usepackage{cite}
\usepackage{amsmath,amssymb,amsfonts}
\usepackage{algorithm}
\usepackage{algorithmic}
\usepackage{graphicx}
\usepackage[T1]{fontenc}
\usepackage{lmodern}
\usepackage{textcomp}
\usepackage{xcolor}
\usepackage{booktabs}
\usepackage{multirow}
\usepackage{url}
\usepackage{balance}
\usepackage{comment}
\def\BibTeX{{\rm B\kern-.05em{\sc i\kern-.025em b}\kern-.08em
    T\kern-.1667em\lower.7ex\hbox{E}\kern-.125emX}}

\newcommand{\wer}{\mathrm{WER}}

\newif\ifblind
\blindtrue
\newcommand{\anon}[2]{\ifblind #2\else #1\fi}
\newcommand{\ouruniv}{\anon{the University of Iceland}{an anonymized university}}

\newcommand{\sdkname}{\anon{the Treble SDK}{an industrial hybrid wave/GA simulator}}

\begin{document}

\title{FFASR: Benchmarking Far-Field Automatic Speech Recognition using High-Fidelity Simulated RIRs}

\author{
\IEEEauthorblockN{\small Shivam Saini\IEEEauthorrefmark{1}, Eric Bezzam\IEEEauthorrefmark{2}, Georg Götz\IEEEauthorrefmark{1}, Alessia Milo\IEEEauthorrefmark{1}, Steinar Guðjónsson\IEEEauthorrefmark{1},\\
\small Konstantinos Gkanos\IEEEauthorrefmark{1}, Finnur Pind\IEEEauthorrefmark{1}, Daniel Gert Nielsen\IEEEauthorrefmark{1}}
\IEEEauthorblockA{\IEEEauthorrefmark{1}\small \textit{Treble Technologies, Reykjav\'{\i}k, Iceland}\\
\IEEEauthorblockA{\IEEEauthorrefmark{2}\small \textit{Hugging Face, Paris, France}
\\
*Correspondence: shivam@treble.tech}}
}

\maketitle

\begin{abstract}
Far-field automatic speech recognition (ASR) degrades under reverberation, noise, and talker motion, yet the benchmarks that drive model selection emphasize close-microphone speech.
We present \emph{FFASR}, a held-out corpus of 15{,}637 utterances and an open leaderboard spanning nine conditions, each varying a single acoustic factor: anechoic near-field speech, a measured-versus-simulated office-lab pair, static far-field mixtures at high/mid/low signal-to-noise ratio (SNR), and moving-talker variants at matched SNR.
Dry speech from 15 talkers is convolved with hybrid wave/geometrical-acoustics room impulse responses from 14 furnished rooms; because the speech is newly recorded and the test waveforms are never released, the corpus resists training-data contamination.
Across contemporary systems, mean word error rate (WER) rises from 4.4\% near-field to 41.3\% in the static low-SNR condition; a moving talker adds a small but consistent penalty at matched SNR; and on the office-lab pair, measured and simulated WER agree to within about 1.7\,pp on average.
These results support high-fidelity simulation as a scalable proxy for measured far-field evaluation under the conditions we test.
\end{abstract}


\section{Introduction}
\label{sec:intro}

Consumer and enterprise voice interfaces increasingly operate at meter-scale distances from the microphone: smart speakers, conference systems, and wearable assistants capture speech only after room reflections, diffraction, and competing noise have reshaped the waveform.
The resulting mismatch is well documented, reverberation smears phonetic cues over hundreds of milliseconds while additive noise suppresses low-energy consonants, and has motivated a long line of dedicated evaluations, from the REVERB challenge~\cite{kinoshita2016reverb} to the CHiME series~\cite{barker2018chime5,watanabe2020chime6}, AMI~\cite{carletta2005ami}, and VOiCES~\cite{richey2018voices}.
Yet the benchmarks that drive day-to-day model selection remain dominated by close-microphone read speech such as LibriSpeech~\cite{panayotov2015librispeech} and its derivatives~\cite{srivastav2025openasr}, and modern foundation models are routinely \emph{trained} on simulated reverberant data~\cite{ko2017augmentation,kim2017interspeech} without a corresponding standardized far-field \emph{test}.

\begin{figure}[!t]
  \includegraphics[width=\linewidth]{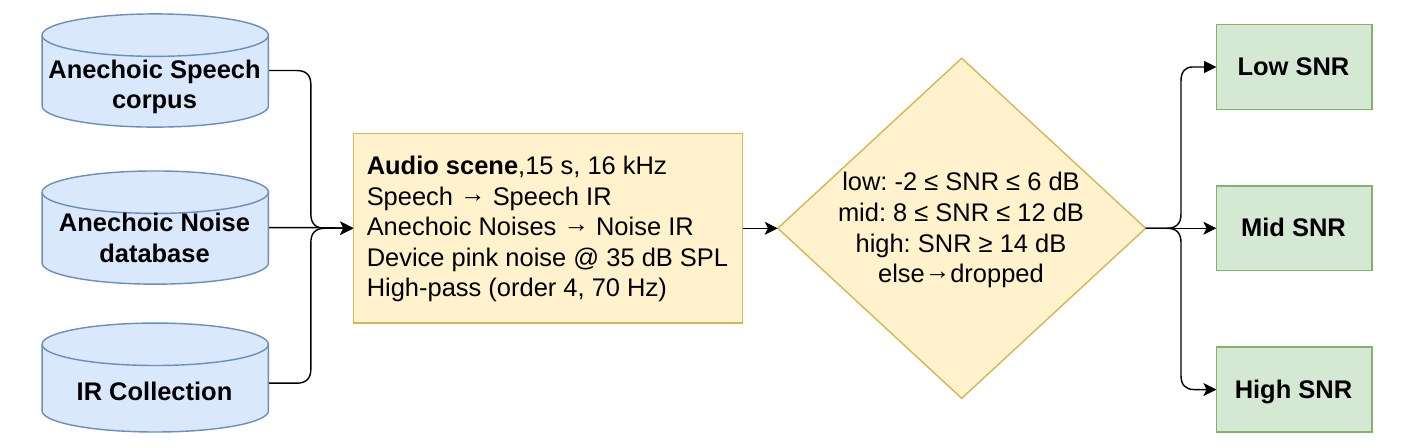}
  \caption{Data-generation pipeline. Newly recorded anechoic speech is convolved with hybrid wave/GA room impulse responses and seeded noise draws; each mixture is labelled by its post-render SNR.}
  \label{fig:pipeline}
\end{figure}

Constructing such a test set is a balancing act: it needs physical realism, broad and controllable acoustic coverage, and source material novel enough to resist training-data contamination, and these goals pull against one another.
Measured corpora~\cite{eaton2016ace,szoke2019but,hadad2014multichannel} offer physical ground truth but cover few geometries and fixed SNR statistics, and once they have been public for several years their audio and transcripts leak into web-scale training crawls, raising contamination risk.
Classical simulators based on geometrical acoustics (GA), such as the image-source method~\cite{allen1979image, scheibler2018pyroomacoustics,diazguerra2021gpurir}, scale arbitrarily but omit diffraction, scattering, and low-frequency modal behavior, exactly the wave phenomena that dominate below the Schroeder frequency of typical rooms.
Hybrid engines that couple wave-based solvers~\cite{pind2019spectral} with GA above a crossover frequency have narrowed this realism gap~\cite{treblesdk,mullins2025treble10,tang2022gwa,chen2022soundspaces}, and recent hybrid higher-order Ambisonics RIR datasets \cite{mullins2025treble10,saini2026hifi} demonstrated that such room impulse responses (RIRs) support far-field ASR research at scale.
In our own preliminary leaderboard runs, several systems that were almost indistinguishable on near-field speech separated sharply once the same utterances were rendered in rooms, which is what convinced us a dedicated far-field test set was needed.
No existing resource combines all of this: a fixed far-field test set whose audio stays private, separate SNR and source-motion factors, a measured anchor for the sim-to-real gap, and shared infrastructure that scores different model families under identical decoding and normalization rules. \textbf{FFASR} is built to close that gap; Fig.~\ref{fig:pipeline} gives a one-view overview and Table~\ref{tab:conditions} lists the nine conditions. We contribute:

  \textbf{A benchmark and open infrastructure.} A nine-condition far-field corpus (Table~\ref{tab:conditions}) built from newly recorded anechoic speech (889 utterances, 15 talkers) convolved with hybrid wave/GA RIRs from 14 furnished rooms (53--367\,m$^3$), plus a measured office-lab triplet for sim-to-real calibration. A deterministic, seeded pipeline shares transcripts, talkers, and noise draws across conditions, so per-condition WER differences come only from the acoustic factor being varied. An open leaderboard scores each submission in an isolated GPU sandbox~\cite{hfjobs2024} and never releases the test waveforms (Sec.~\ref{sec:benchmark}--\ref{sec:eval}).

  \textbf{Empirical findings} from evaluating contemporary systems, from compact CTC models to large speech LLMs (Sec.~\ref{sec:results}): clean-speech rankings hide large far-field differences, separating systems that look equivalent on clean speech and triggering insertion-dominated hallucination at low SNR; source motion adds a smaller additional cost at matched SNR; and the measured and simulated office-lab renders give similar WER across architectures, suggesting the simulated renders are close enough to measured ones for this evaluation.

\begin{table}[b]
  \centering
  \caption{FFASR evaluation conditions. $N$ is the number of clips in each \emph{packed} split.}
  \label{tab:conditions}
  \begin{tabular}{@{}lrl@{}}
    \toprule
    Condition & $N$ & Description \\
    \midrule
    Near field & 889 & Dry anechoic speech \\
    Lab simulated & 2000 & Stage~A hybrid render \\
    Lab measured & 2000 & Stage~A measured mono \\
    High SNR & 1746 & Static, SNR $\geq 14$\,dB \\
    Mid SNR & 1938 & Static, 8--12\,dB \\
    Low SNR & 1647 & Static, $\leq 6$\,dB \\
    Moving high & 1808 & Trajectory + high SNR \\
    Moving mid & 1816 & Trajectory + mid SNR \\
    Moving low & 1793 & Trajectory + low SNR \\
    \bottomrule
  \end{tabular}
\end{table}

\section{Related Work}
\label{sec:related}

\subsection{Far-Field Corpora and Challenges}
The REVERB challenge~\cite{kinoshita2016reverb} established reverberant evaluation with simulated and measured single- and eight-channel data. CHiME-5/6~\cite{barker2018chime5,watanabe2020chime6} moved to unsegmented multi-talker dinner parties, while AMI~\cite{carletta2005ami} and LibriCSS~\cite{chen2020libricss} target meetings and overlapped speech.
These corpora combine many degradation factors at once, such as overlap, disfluency, and distant arrays, which makes them excellent integration tests but poor instruments for isolating a single far-field factor; convolution with measured RIRs is controllable but static.
FFASR complements these with single-talker, factorized conditions with matched lexical content across all nine splits.

\subsection{Simulation for Far-Field Speech Processing}
Simulated RIRs are a standard tool for \emph{training} robust ASR~\cite{ko2017augmentation,kim2017interspeech} and for privacy-preserving on-device development~\cite{khan2025novel}; our concern here is instead their fidelity for \emph{evaluation}.
The hybrid wave/GA rendering in \sdkname  (Sec.~\ref{sec:simulation}) has been validated against measurement across several tasks~\cite{trebleauralizer2023,gotz2025room,wavebasedframework2024,smallroom2024,mullins2025treble10,goetz2026enhancement,alessiaiwaenc}.
Most relevant to the present evaluation, \cite{gotz2025room} investigates the simulation-to-real gap when evaluating audio algorithms on simulated data with varying fidelity rather than measurements, while \cite{goetz2026enhancement} and \cite{alessiaiwaenc} report that the added physical fidelity of hybrid RIR training data over image-source augmentation transfers to a downstream ASR metric, reducing median WER on \emph{measured} test data by up to 38\% relative.
FFASR builds on this line, using 14 furnished scenes with many source--receiver positions and directional noise sources.

\subsection{Robust ASR Systems}
The systems we evaluate are drawn from across the current architecture space: web-scale weakly supervised encoder--decoders (Whisper~\cite{radford2023whisper}) and compressed variants~\cite{gandhi2023distilwhisper,kamahori2025liteasr}; Conformer~\cite{gulati2020conformer}/FastConformer~\cite{rekesh2023fastconformer} encoders with CTC or transducer~\cite{graves2012sequence} decoders, including the token-and-duration transducer (TDT)~\cite{xu2023tdt} used by Parakeet; Canary~\cite{puvvada2024canary}; OWSM~\cite{peng2025owsmv4,peng2024owsmctc}; speech-aware LLMs (Granite Speech~\cite{saon2025granite}); wav2vec~2.0~\cite{baevski2020wav2vec} CTC baselines via SpeechBrain~\cite{ravanelli2024speechbrain}; and streaming-oriented compact models (Moonshine~\cite{jeffries2024moonshine}). We treat these as the systems under test rather than as related methods, and report their behavior in Sec.~\ref{sec:results}.

\subsection{Leaderboards and Evaluation Methodology}
The Open ASR Leaderboard~\cite{srivastav2025openasr} standardized text normalization and RTFx reporting for predominantly close-talk English test sets and documented how normalization choices alone shift WER rankings.
FFASR adopts the same normalizer and efficiency metric so that scores are directly comparable, while adding the far-field axes those suites lack.
Prior work also documents Whisper-family hallucination on degraded or silent input~\cite{koenecke2024careless}; our low-SNR conditions elicit this failure mode systematically and expose it as WER exceeding 100\% (Sec.~\ref{sec:rankings}).

\section{The FFASR Benchmark}
\label{sec:benchmark}

Fig.~\ref{fig:pipeline} summarizes the generation flow and Algorithm~\ref{alg:gen} states it step by step.
A few deliberate choices shape the corpus.
All source speech is newly recorded and the test waveforms are never distributed, so a model cannot score well simply by having seen the test data in training.
Each condition changes a single acoustic factor while holding transcripts, talkers, and seeded noise draws fixed, so a WER gap between two conditions can be attributed to that factor rather than to a change in content.
And rendering uses a hybrid wave/GA solver that has been checked against measurement, which keeps the simulated rooms physically grounded rather than idealized.
The nine resulting conditions are listed in Table~\ref{tab:conditions}.

\begin{algorithm}[t]
  \caption{FFASR scene generation (one mixture)}
  \label{alg:gen}
  \small
  \begin{algorithmic}[1]
    \REQUIRE one dry utterance $s$, a pool of anechoic interferers, the RIR collections, and a shared random seed
    \STATE draw $1$--$2$ interferers $\{n_k\}$ using the shared seed, so the same draw recurs across conditions
    \STATE pick a room and the target/receiver positions, and load the target's RIR $h_{\mathrm{s}}$ and one RIR $h_k$ per interferer
    \STATE place each source in the room by convolution: target $y_{\mathrm{s}} = h_{\mathrm{s}} \ast s$, background $y_n = \sum_k h_k \ast n_k$ \COMMENT{time-varying $h_{\mathrm{s}}(t,\tau)$ if the target moves}
    \STATE pass every render through the device front-end: a $70$\,Hz high-pass plus added pink self-noise $d(t)$
    \STATE form the mixture $y = y_{\mathrm{s}} + y_n + d$, and mark the speech-active region $\mathcal{A}$ of $y_{\mathrm{s}}$ with an ITU-T~P.56 detector
    \STATE measure the SNR of $y$ over $\mathcal{A}$ using \eqref{eq:snr}
    \IF{the SNR lands in the High, Mid, or Low band}
      \STATE keep $y$ and label it with that band
    \ELSE
      \STATE discard $y$: it fell in a guard band between bands
    \ENDIF
    \ENSURE mono $16$\,kHz mixture $y$ with its SNR-band label
  \end{algorithmic}
\end{algorithm}

\subsection{Anechoic Source Speech}
\label{sec:anechoic}

Clean speech is recorded in the anechoic chamber at \ouruniv.
Fifteen subjects (6 female, 9 male; \anon{all Treble employees or their spouses}{who gave informed consent for research use of the recordings}, who gave informed consent for research use of the recordings) each read 80 English sentences from a fixed prompt list.
Fifteen talkers is a compromise: enough to average over individual voice and accent, but small enough that every talker appears in every condition, so no split is confounded by a change in voices.
Two talkers are native English speakers and thirteen are non-native, spanning different native-language backgrounds, so the pool reflects the accent diversity of deployed voice interfaces.
FFASR isolates the \emph{far-field acoustic} factor rather than accent: all nine conditions share this single talker pool, so each model's own near-field score is its clean-speech reference. Accent thus varies across talkers but is held constant across conditions (Sec.~\ref{sec:limitations}).
Of the 1{,}200 prompt recordings (15 talkers $\times$ 80 prompts), a single annotator screened every recording and discarded 311 (26\%) for mispronunciations, disfluencies, truncated prompts, or audible recording defects, leaving \textbf{889} utterances.
Because no public read-speech test set is reused, leaderboard scores reflect acoustic generalization rather than memorized lexicon or speaker profiles.
Each utterance is normalized and rendered at 60\,dB SPL free-field level.

\subsection{Hybrid Room Acoustic Simulation}
\label{sec:simulation}

Let $s(t)$ denote the dry target utterance, $\{n_k(t)\}_{k=1}^{K}$ the anechoic interferer signals, and $h_{\mathrm{s}}, h_k$ the RIRs to the receiver from the target speaker (subscript $\mathrm{s}$) and the $k$-th noise source.
Each scene uses up to $K \le 2$ interferers, and each static far-field mixture is
\begin{equation}
  y(t) = (h_{\mathrm{s}} \ast s)(t) + \sum_{k=1}^{K}(h_k \ast n_k)(t) + d(t),
  \label{eq:mixture}
\end{equation}
where $\ast$ denotes convolution and $d(t)$ is device self-noise.

RIRs are drawn from three precomputed collections:
\begin{itemize}
  \item \textbf{Static Sources}: 14 scenes spanning living spaces, offices, meeting rooms, classrooms, and restaurants, with volumes 53--367\,m$^3$.
  \item \textbf{Moving Sources}: paired moving variants of the same rooms with time-varying source paths.
  \item \textbf{Measured-Simulated pairs}: 44 measured RIRs and their simulated counterparts over six receivers in a single office, in three configurations that change only the acoustic treatment: bare brick walls($\mathrm{RT}_{60}\approx0.9$\,s), the same room furnished ($\approx0.6$\,s), and the room fitted with wall absorber panels ($\approx0.3$\,s).
\end{itemize}

All furnished-room IRs use \emph{hybrid} simulation in \sdkname~\cite{treblesdk}: a wave-based discontinuous Galerkin finite-element solver~\cite{pind2019spectral} up to a 2\,kHz crossover and GA above, computed at 32\,kHz and stored as 8th-order ambisonics (81 channels) before rendering to mono 16\,kHz.
The 2\,kHz crossover sits below most consonantal energy (2--8\,kHz), so the wave solver mainly improves the low- and mid-frequency modal and diffraction behavior.
The external validations above~\cite{mullins2025treble10,goetz2026enhancement} and the measured anchor (Sec.~\ref{sec:simreal}) confirm that this added fidelity still transfers to ASR-level WER.
Receivers are restricted to heights $1.0 < z < 2.3$\,m, consistent with tabletop and wall-mounted devices; elevated sources ($z > 1.9$\,m) model HVAC-like emitters.
Each scene places one directive target talker and up to two noise paths: a mixture pool and static-like ambient noises from AID~\cite{gotz2022aid}.

\subsection{Noise, SNR Ranges, and Device Chain}
\label{sec:snr}

Reverberant energy builds up differently depending on room geometry and absorption, so the source-level ratio set in advance is a poor predictor of the SNR observed at the receiver.
We therefore measure SNR \emph{after} rendering.
A fixed device chain is first applied at the listener to emulate the analog front-end of consumer hardware: a 4th-order Butterworth high-pass at 70\,Hz and pink device self-noise $d(t)$ at 35--40\,dB SPL.
The 70\,Hz corner approximates the low-frequency rolloff of the small transducers in smart speakers and phones and removes sub-band rumble those devices would not capture, so the signal reaching the recognizer matches deployed hardware rather than an idealized full-band render.
The SNR is then the energy ratio between the target render $y_{\mathrm{s}} = h_{\mathrm{s}} \ast s$ and the full background, namely the interferer render $y_n = \sum_k h_k \ast n_k$ plus the device noise $d(t)$.
Both energy terms are accumulated only over the speech-active region $\mathcal{A}$ of the target render $y_{\mathrm{s}}$, where $\mathcal{A}$ is the set of samples lying in frames flagged active by an ITU-T~P.56 detector~\cite{itu_p56}, so that silent and inter-word pauses do not contribute:
\begin{equation}
  \mathrm{SNR} = 10\log_{10}\frac{\sum_{t\in\mathcal{A}} y_{\mathrm{s}}^2(t)}{\sum_{t\in\mathcal{A}} \big(y_n(t)+d(t)\big)^2}.
  \label{eq:snr}
\end{equation}
Keeping $d(t)$ in the denominator means the label reflects the SNR actually presented to the recognizer, not an idealized speech-to-interferer ratio.
Each mixture is assigned to one of three SNR ranges, Low $[-2,6]$\,dB, Mid $[8,12]$\,dB, and High $\geq 14$\,dB.
The bands are chosen to bracket qualitatively different listening regimes, adverse, effortful, and comfortable, rather than to trace a fine SNR curve, which keeps the conditions interpretable and the per-condition splits large.
The transition bands $(6,8)$ and $(12,14)$\,dB act as guard bands and are excluded, so labels stay unambiguous under small changes in the render.
For reproducibility, a shared integer seed fixes the interferer draw, room, and source/receiver positions so any scene can be regenerated, mixtures are peak-limited below 0\,dBFS to prevent clipping before \eqref{eq:snr} is evaluated, and all renders are written as 16\,kHz mono WAV.

\subsection{Moving Talker Conditions}
\label{sec:moving}

To simulate a moving talker, the static target RIR is replaced by a time-varying RIR that evolves along a predefined source trajectory, while keeping the interferers fixed. Therefore, the static convolution in \eqref{eq:mixture} becomes a time-varying convolution,
\begin{equation}
  y(t) = \sum_{\tau} h_{\mathrm{s}}(t,\tau)\, s(t-\tau) + \sum_{k=1}^{K}(h_k \ast n_k)(t) + d(t),
  \label{eq:tv}
\end{equation}
where $\tau$ is the discrete time lag and $h_{\mathrm{s}}(t,\tau)$ is the target impulse response at time $t$, interpolated along the source trajectory; the interferers stay static.
We synthesize $h_{\mathrm{s}}(t,\tau)$ with the frequency-domain interpolation method of Mullins and Sampedro Llopis~\cite{mullins2026motion}.
That work compared three offline movement-simulation algorithms. Frequency-domain interpolation achieved the most accurate interaural-time-difference reproduction and the highest perceptual (MUSHRA) similarity to real moving-source recordings, outperforming time-domain crossfading and nearest-neighbor switching.
We reuse the same talker and interferer signals as in the static scenes. The static target RIR is replaced by a trajectory that traverses the complete moving-IR path over 6\,s of speech within a 15 s scene, corresponding to an effective spatial update rate of $\approx$15.6\,Hz.We define Moving High, Mid, and Low using the same SNR ranges as the corresponding static conditions; the labels refer to SNR rather than movement speed.

The moving and static splits are matched by SNR range but not perfectly seed-aligned: because SNR is measured after rendering and motion changes the source--receiver distance within an utterance, a seed in one static range can land in an adjacent range once it moves, so a naive moving-minus-static difference mixes the motion effect with a small reshuffling across ranges.
At the scale of FFASR this is minor: each moving range holds $\approx$1{,}800 utterances from the same talker, room, and noise seeds as its static counterpart, the guard bands limit boundary leakage, and the matched splits' post-render SNR distributions stay closely aligned.
The aggregate comparison is therefore stable enough for the leaderboard-level analysis reported here, and the motion penalty is consistently positive across competitive models and SNR ranges (Sec.~\ref{sec:results}).

\subsection{Dataset Statistics and Splits}
\label{sec:stats}

\begin{figure}[!t]
  \centering
  \includegraphics[width=0.96\linewidth]{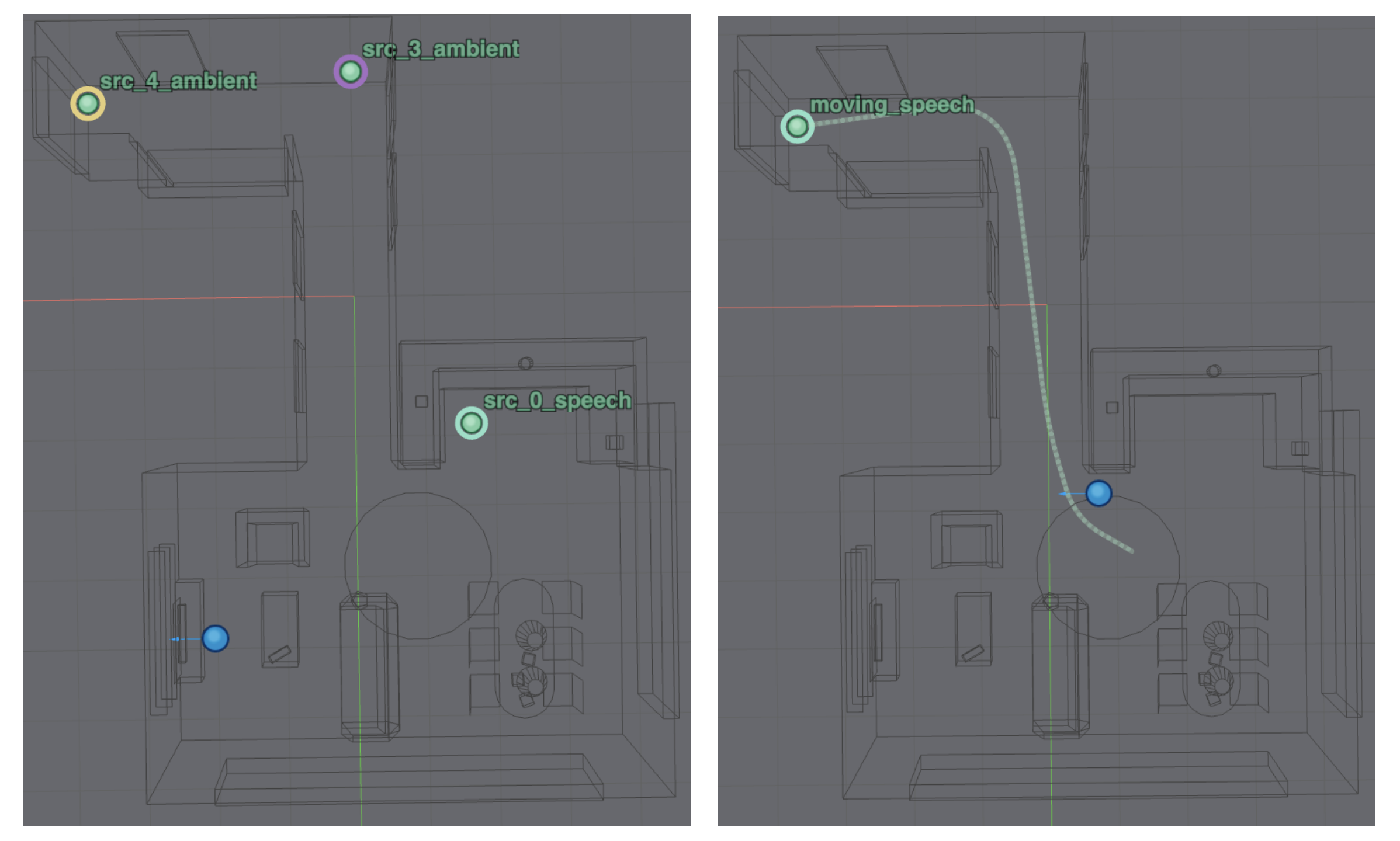}
  \caption{Example scene from the dataset. Left: the target speech source, the directional transient and ambient noise sources, and the receiver. Right: an example trajectory of the moving speech source.}
  \label{fig:scene}
\end{figure}

Table~\ref{tab:conditions} lists the packed evaluation splits and Table~\ref{tab:acoustic} summarizes the acoustic statistics of the furnished-room renders.
Up to 2000 samples per condition are kept by subsampling that keeps the 15 talkers balanced; the anechoic split keeps all 889 dry utterances.
In total 15{,}637 clips are drawn from a pool of 17{,}700 renders, and a full nine-condition run transcribes $\approx$54.9\,h of audio.
The talkers contribute roughly evenly (per-talker counts 821--992).
At the clip level the corpus is about 60\% male / 40\% female and 86\% non-native / 14\% native, tracking the 9:6 and 13:2 splits of the talker pool.
The renders span the 14 rooms of Sec.~\ref{sec:simulation} and more than 40 receiver positions (seated and standing listeners, wall-, shelf-, and counter-mounted devices), with post-render SNR roughly balanced across the three ranges.
Fig.~\ref{fig:scene} shows an example scene.

\begin{table}[t]
  \centering
  \caption{Acoustic statistics of the FFASR furnished-room renders.}
  \label{tab:acoustic}
  \footnotesize
  \setlength{\tabcolsep}{6pt}
  \begin{tabular}{@{}lrrrr@{}}
    \toprule
    Quantity & Min & Max & Mean & SD \\
    \midrule
    Room volume (m$^3$)            & 17.47 & 367.41 & 128.52 & 77.26 \\
    $\mathrm{T}_{20}$ (s)         & 0.19 & 1.29 & 0.60 & 0.3 \\
    Source--receiver dist.\ (m)    & 0.3 & 13.8 & 3.98 & 2.2 \\
    $C_{50}$ (dB)                  & -7.8 & 19.61 & 5.3 & 4.5 \\
    Post-render SNR (dB)           & -1.9 & 88.9 & 10.5 & 6.4 \\
    \bottomrule
  \end{tabular}
\end{table}

\section{Evaluation Protocol and Metrics}
\label{sec:eval}

\subsection{Leaderboard Infrastructure}
Submissions arrive through a public Hugging Face Space~\cite{ffasr2026}.
Each job clones the evaluation harness inside an isolated UV sandbox on an NVIDIA L4 GPU~\cite{hfjobs2024}, loads the user-specified checkpoint, and writes a versioned JSON artifact to a private bucket; the orchestrating process never imports the inference stack, avoiding dependency collisions across model families.
The packed test waveforms are never exposed to submitters, which is the mechanism that resists contamination.
For the same reason we do not accept closed, API-only systems, since scoring them would require sending the held-out audio to a third party.

\begin{table}[!t]
  \centering
  \caption{Distribution of WER (\%) across the 15 leading evaluated systems, summarized per condition. The lower panel reports the two derived effects in percentage points (pp).}
  \label{tab:main}
  \footnotesize
  \setlength{\tabcolsep}{5pt}
  \begin{tabular}{@{}lrrrrr@{}}
    \toprule
    Condition & Mean & Med & SD & Min & Max \\
    \midrule
    Near field        & 4.4  & 4.3  & 0.4 & 3.8  & 5.4 \\
    Lab measured      & 29.0 & 27.8 & 6.7 & 20.0 & 45.0 \\
    Lab simulated     & 27.7 & 26.9 & 6.9 & 18.6 & 42.6 \\
    Static High SNR   & 10.1 & 10.2 & 2.2 & 6.7  & 15.1 \\
    Static Mid SNR    & 20.5 & 20.3 & 4.7 & 13.9 & 29.5 \\
    Static Low SNR    & 41.3 & 40.2 & 8.3 & 28.4 & 57.0 \\
    Moving High SNR   & 11.5 & 11.5 & 2.3 & 8.2  & 16.8 \\
    Moving Mid SNR    & 23.2 & 23.5 & 5.1 & 15.3 & 32.3 \\
    Moving Low SNR    & 43.5 & 42.4 & 8.6 & 30.7 & 60.0 \\
    \midrule
    \multicolumn{6}{@{}l}{\emph{Derived effects (pp)}}\\
    Sim2real $\Delta$ (meas-sim) & 1.3 & 1.3 & 1.2 & $-2.5$ & 2.8 \\
    Motion penalty, High SNR       & 1.3 & 1.5 & 0.5 & 0.1 & 1.9 \\
    Motion penalty, Mid SNR        & 2.7 & 2.6 & 0.9 & 1.2 & 4.7 \\
    Motion penalty, Low SNR        & 2.2 & 2.3 & 1.0 & 0.1 & 3.8 \\
    \bottomrule
  \end{tabular}
\end{table}

\subsection{Decoding and Metrics}
Every utterance is transcribed at 16\,kHz with batch size~1.
References and hypotheses pass through the Whisper \texttt{EnglishTextNormalizer}~\cite{radford2023whisper} (lowercasing, punctuation removal, contraction expansion) before scoring, following the Open ASR Leaderboard protocol~\cite{srivastav2025openasr}.
WER is the length-normalized Levenshtein cost
\begin{equation}
  \wer = \frac{S + D + I}{N},
  \label{eq:wer}
\end{equation}
with substitutions $S$, deletions $D$, insertions $I$, and reference length $N$~\cite{morris2004wer}. $\wer$ can exceed 1 when insertions dominate; rather than clipping these values, we report them because they are diagnostic of hallucination~\cite{koenecke2024careless}.

The primary ranking metric is \textbf{Average WER} over four core scenarios: near-field speech plus the three static SNR ranges.
This ranking should be read together with the per-condition results on the live leaderboard, since changing which conditions enter the average will in general change the ordering.
Throughput is reported as $\mathrm{RTFx}$ batch size~1 on the reference GPU, matching~\cite{srivastav2025openasr}.

\begin{figure*}
  \centering
  \includegraphics[width=0.9\linewidth]{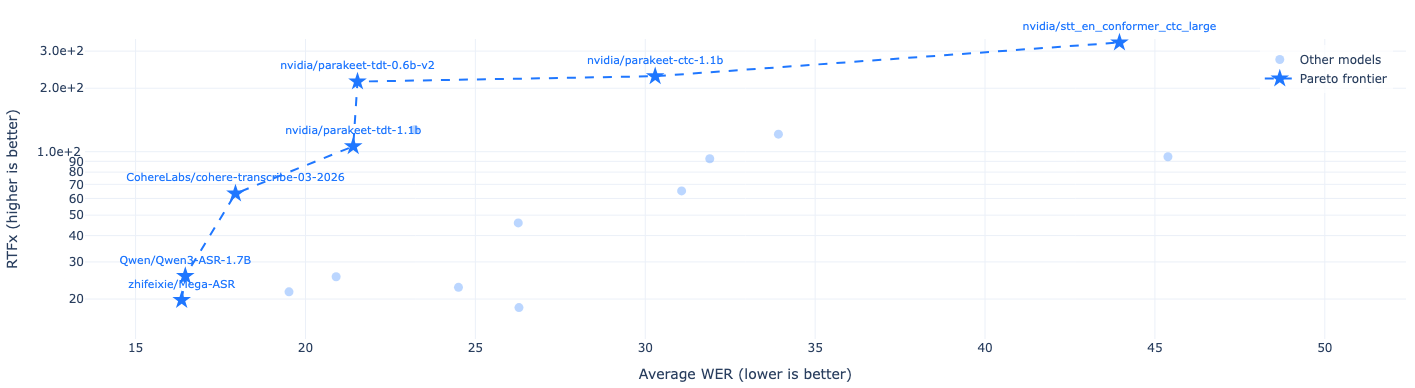}
  \caption{Average WER vs.\ RTFx on NVIDIA L4 (batch size~1). Efficient CTC/TDT models occupy the upper-left region.}
  \label{fig:pareto}
\end{figure*}

\section{Results and Discussion}
\label{sec:results}

At the time of submission the leaderboard holds 24 publicly released systems spanning CTC, RNN-T/TDT, encoder--decoder, and speech-LLM architectures, all scored under the same harness, normalizer, and decoding settings.
We summarize the field through the per-condition distribution in Table~\ref{tab:main} and report what we find across conditions rather than a ranking of named systems, since that ranking will change as new models appear; model identities and their efficiency are shown in Fig.~\ref{fig:pareto}.
Where we quantify an effect below, we give its mean across the 15 systems together with a 95\% interval, $\text{mean}\pm1.96\,\mathrm{SD}/\sqrt{15}$, that reflects how tightly the across-system mean is pinned down by our sample of systems (not a per-utterance bootstrap, which would need the seed-paired data deferred to a future release, Sec.~\ref{sec:limitations}).

\subsection{Far-field speech is much harder than near-field}
\label{sec:rankings}
Near-field WER is uniformly low and tightly clustered (mean 4.4\%, all systems within 3.8--5.4\%), so clean speech barely separates the field.
Reverberation and noise change this: mean WER rises to 10.1\% at high SNR, 20.5\% at mid SNR, and 41.3\% at low SNR, and the spread across systems widens from under 2\,pp near-field to roughly 30\,pp at low SNR.
The largest single step is from mid to low SNR ($\approx$21\,pp in the mean).
We read this as the regime where the noise floor approaches speech energy and masks low-energy consonant cues, though we infer this from the error pattern rather than measure it directly.
The degradation is also architecture-dependent.
Aggressively compressed encoder--decoders (distilled and ``lite'' Whisper variants, and the most compact Canary checkpoints) match the rest of the field near-field yet fall furthest behind at low SNR, while large speech-LLM and well-trained transducer systems degrade more gracefully.
For the checkpoints we evaluate this suggests that compression trades away reverberation robustness first; we cannot rule out that better-trained small models would behave differently.
The most extreme case is outright hallucination: at low SNR the smallest Whisper checkpoints exceed 100\% WER (\texttt{whisper-base} 117\%, \texttt{whisper-tiny} 137\%)~\cite{radford2023whisper} even though the same checkpoints score below 10\% on clean speech~\cite{srivastav2025openasr}.
Inspecting these outputs, the errors are insertion-dominated: during noise-only or low-energy segments the model emits repeated phrases or fabricated sentences~\cite{koenecke2024careless}, so the hypothesis runs far longer than the reference and WER passes 100\%.
Far-field WER therefore separates robust from brittle systems where clean-speech WER cannot.

\subsection{Source motion adds a consistent penalty}
\label{sec:motion}
Adding source motion at matched SNR raises WER across the board: the static-to-moving penalty (Table~\ref{tab:main}, lower panel) is positive for essentially every competitive system in every SNR range.
The penalty is not uniform with SNR. It is largest at mid SNR (mean $+2.7$\,pp, 95\% interval $[2.2,3.2]$), smaller at low SNR ($+2.2$\,pp, $[1.7,2.7]$), and smallest at high SNR ($+1.3$\,pp, $[1.0,1.6]$); all three intervals exclude zero.
We had expected the penalty to grow as SNR fell, tracking the overall error rate. Instead the largest average penalty sits at mid SNR.
A plausible reading is masking: at low SNR additive noise already dominates the errors and hides the kinematic contribution, while at high SNR there is little accuracy left to lose, leaving the mid-SNR regime, where the within-utterance change in direct-to-reverberant ratio is the main remaining difficulty, as the place a moving talker costs most. We offer this as an interpretation, not a measured mechanism.
Because the moving conditions are released as a beta and are matched by SNR range rather than seed-paired (Sec.~\ref{sec:moving}), we report these as an aggregate trend rather than a per-model paired estimate.

\subsection{High-fidelity simulation matches measurement}
\label{sec:simreal}
On the matched office-lab pair, where only the rendering engine differs, measured and simulated WER agree closely: across systems the mean absolute difference is 1.7\,pp and the mean signed difference is $+1.3$\,pp (95\% interval $[0.7,1.9]$).
The direction is not consistent, measured is marginally harder for most systems but easier for some, so the residual gap behaves like small, roughly zero-mean differences between two realistic renderings rather than a systematic bias of the simulator.
The gap is also model-dependent: robust systems sit within about 1\,pp, while brittle ones amplify small statistical differences between the measured and simulated RIRs into large WER differences (up to $+12.7$\,pp), which is why we report it per system rather than as a single constant.
The agreement is not from one easy operating point: the anchor spans three configurations ($\mathrm{RT}_{60}\approx 0.3$--$0.9$\,s) and 44 RIR pairs over six receivers, and it reproduces on a larger model set the downstream-ASR fidelity reported for wave-based/hybrid rendering~\cite{gotz2025room}.
Because the anchor is a single office lab (Sec.~\ref{sec:limitations}), we read this as site-specific evidence: within those bounds, the simulated office-lab renders are close enough to the measured ones for this evaluation.

\subsection{Speed-accuracy trade-off}
Fig.~\ref{fig:pareto} plots Average WER against RTFx and shows a clear accuracy--throughput tension.
The most accurate systems are large and comparatively slow, anchoring the low-WER but not the high-throughput end of the Pareto frontier.
The efficient end is dominated by frame-synchronous transducer and CTC decoders~\cite{xu2023tdt,rekesh2023fastconformer}, which reach competitive WER at one to two orders of magnitude higher throughput, whereas autoregressive speech-LLMs fall well below real-time.
The frontier reads as a deployment guide: transducer/CTC models for latency-bound use, attention-decoder or speech-LLM systems when accuracy is paramount.

\section{Limitations and Future Work}
\label{sec:limitations}

\textbf{Mono input.} FFASR currently scores single-channel renders even though the underlying RIRs are 8th-order ambisonic. We chose mono because all 24 evaluated systems accept only single-channel input, so a mono harness is what makes them directly comparable; the cost is that beamforming and array front-ends are not yet exercised.

\textbf{Measured-room coverage.} The measured anchor is one office, rendered in three configurations spanning $\mathrm{RT}_{60}\approx 0.3$--$0.9$\,s (absorber panels, furnishing, bare brick). This covers a wide range of reverberation at a single site, so the sim-to-real agreement we report should not be read as a general bound. We used the rooms we had measured pairs for; extending to more measured rooms is the clearest next step.

\textbf{Speaker and accent coverage.} Source material is English read speech from 15 talkers. Accent varies across talkers but is held constant across conditions by design, which is what lets each near-field score serve as a clean reference; the trade-off is that FFASR measures far-field robustness, not accent robustness, and broader demographic coverage would need a larger recording effort.

\textbf{Read vs.\ conversational speech.} The prompts are read sentences. This keeps lexical content matched across conditions, but omits the disfluency, overlap, and turn-taking of spontaneous speech, which corpora such as CHiME~\cite{watanabe2020chime6} and AMI~\cite{carletta2005ami} target directly.

\textbf{Moving-source beta.} The moving conditions are a beta. The splits are matched by SNR range rather than seed-paired, and the rendering may still be revised, so we report the motion penalty as an aggregate trend rather than a per-model paired estimate.

\textbf{Future work.} Planned extensions include multichannel and multi-speaker input that uses the existing ambisonic RIRs, moving toward joint ASR--diarization evaluation~\cite{watanabe2020chime6,chen2020libricss}; more measured rooms; a seed-paired motion metric with confidence intervals; and streaming latency tiers that report word-emission delay alongside RTFx~\cite{jeffries2024moonshine}.

\section{Conclusion}
\label{sec:conclusion}

FFASR is a far-field ASR benchmark whose source speech is newly recorded and whose test audio stays private, built so that a single acoustic factor changes between conditions. The practical aim of FFASR is to make far-field robustness visible in comparisons that look saturated on near-field speech.
Two patterns stand out from the current field. Clean-speech WER has saturated, the systems we evaluate sit within 2\,pp of each other near-field, yet the same systems spread by more than 20\,pp once reverberation and noise are added, and the smallest models tip into insertion-dominated hallucination at low SNR. Moreover, on the office-lab anchor, measured and simulated WER track each other to within about 1.7\,pp across architectures, close enough for simulation to stand in for measurement at that site.



\balance
\vfill
\bibliographystyle{IEEEtran}
\bibliography{refs}

\end{document}